\documentclass[apj]{emulateapj}
\usepackage{color}
\usepackage{ulem}
\usepackage{amsmath}
\usepackage{amssymb}
\usepackage{amsthm}
\usepackage{graphicx}
\usepackage{multirow}
\usepackage{bm}
\usepackage{subfigure}

 \setcitestyle{notesep={ }}

\newcommand{\be}{\begin{equation}}
\newcommand{\ee}{\end{equation}}
\newcommand{\ba}{\begin{eqnarray}}
\newcommand{\ea}{\end{eqnarray}}
\newcommand{\no}{\nonumber}
\newcommand{\bi}{\begin{itemize}}
\newcommand{\ei}{\end{itemize}}

\newcommand{\Nbody}{N{\mathrm{-body}}}
\newcommand{\LCDM}{\Lambda{\mathrm{CDM}}}
\newcommand{\omegal}{\Omega_{\mathrm{\Lambda}}}
\newcommand{\omegam}{\Omega_{\mathrm{m}}}
\newcommand{\omegab}{\Omega_{\mathrm{b}}}

\newcommand{\mpch}{h^{-1} \, \mathrm{Mpc}}

\newcommand{\kmsmpc}{\mathrm{km s^{-1} Mpc^{-1}}}

\makeatletter

\shorttitle{Scaling Relations}
\shortauthors{Meng et al.}

\begin{document}
%%%%%%%%%%%%%%%%%%%%%%%%%%%%%%%%%%%%%%%%%%%%%%%%%%%%%%%%%%%%%%%%%%%%%%%%%%%%%%%%%%%%%%%%%%%%%%%%%%%
\title{Verifications of scaling relations useful for the
  intrinsic alignment self-calibration}
%\author{Xian-guang Meng\altaffilmark{1,2}, Yu Yu\altaffilmark{3,4,$\star$},Pengjie Zhang\altaffilmark{3,4,5,6,$\dagger$},Yipeng Jing\altaffilmark{3,4,5,6,*}}
\author{Xian-guang Meng\altaffilmark{1,2}, Yu Yu\altaffilmark{3,4,$\star$},Pengjie Zhang\altaffilmark{3,4,5,6},Yipeng Jing\altaffilmark{3,4,5,6}}
\altaffiltext{1}{Key Laboratory for Research in Galaxies and Cosmology, Shanghai Astronomical Observatory, Chinese Academy of Sciences, 
80 Nandan Road, Shanghai, 200030, People’s Republic of China}
\altaffiltext{2}{University of Chinese Academy of Sciences, 
19A Yuquan Road Beijing, China, 100049, People’s Republic of China}
\altaffiltext{3}{Department of Astronomy, School of Physics and Astronomy, Shanghai
Jiao Tong University, Shanghai, 200240, People’s Republic of China}
\altaffiltext{4}{Shanghai Key Laboratory for Particle Physics and
  Cosmology, People’s Republic of China}
\altaffiltext{5}{IFSA Collaborative Innovation Center, Shanghai Jiao Tong
University, Shanghai 200240, People’s Republic of China}
\altaffiltext{6}{Tsung-Dao Lee Institute, Shanghai 200240, People’s Republic of China}
\altaffiltext{$\star$}{yuyu22@sjtu.edu.cn}
%\altaffiltext{$\dagger$}{zhangpj@sjtu.edu.cn}
%\altaffiltext{*}{ypjing@sjtu.edu.cn}

%%%%%%%%%%%%%%%%%%%%%%%%%%%%%%%%%%%%%%%%%%%%%%%%%%%
%%%%%%%%%%%%%%%%%%%%%%%%%%%%%%%%%%%%%%%%%%%%%%%%%%%
\begin{abstract}
The galaxy intrinsic alignment (IA) is a major challenge of weak lensing cosmology. 
To alleviate this problem, Zhang (2010, MNRAS, 406, L95) proposed  a self-calibration method, independent of IA modeling. 
This proposal relies on several scaling relations between 
two-point clustering of IA and matter/galaxy fields, which were
previously only tested with analytical IA models. 
In this paper, these relations  are tested comprehensively with an
$\Nbody$ simulation of $3072^3$ simulation
particles and boxsize $600\mpch$.  
They are verified at the accuracy level of $\mathcal{O}(1)\%$ over 
angular scales and source redshifts of interest. 
We further confirm that these scaling relations are generic,  
insensitive to halo mass, weighting in defining halo ellipticities, photo-$z$ error,  
and misalignment between galaxy ellipticities and halo ellipticities.  
We also present and verify three new scaling relations on the B-mode IA. 
These results consolidate and complete the theory side of the proposed self-calibration technique. 
\end{abstract}


%%%%%%%%%%%%%%%%%%%%%%%%%%%%%%%%%%%%%%%%%%%%%%%%%%%%%%%%
%%%%%%%%%%%%%%%%%%%%%%%%%%%%%%%%%%%%%%%%%%%%%%%%%%%%%%%%
\section{Introduction}
\label{sec:intro}
%%%%%%%%%%%%%%%%%%%%%%%%%%%%%%%%%%%%%%%%%%%%%%%%%%%
The intrinsic alignment (IA) of galaxies, in particular, the spatially
correlated component of galaxy shapes,  is a major challenge of
weak lensing cosmology \citep[see][for a recent review]{Troxel2015}. 
Depending on galaxy types,  it  may significantly contaminate cosmic shear measurement in several ways. 
For weak lensing two-point correlation or its corresponding power spectrum,  
IA directly induces the so-called II term, arising from the IA auto correlation. 
Furthermore, IA can be spatially correlated with the ambient density field 
and therefore spatially correlated with gravitational lensing. 
This induces the so-called GI term \citep{Hirata2004b}.  
These contaminations have been predicted in various simulations \citep[e.g.][]{Croft2000, Heavens2000, Jing2002, 
Heymans2006b, Joachimi2013, Hilbert2017, Xia2017, Wei2018}
and analytical modeling \citep[e.g.][]{Catelan2001, Crittenden2001,  Lee2001, Hirata2004b, Bridle2007, 
Hui2008, Schneider2010, Blazek2011, Blazek2015, Blazek2017,  Joachimi2011, Tugendhat2018}.  
Furthermore, they  have been detected in observations \citep{Brown2002, Hirata2004a, Mandelbaum2006, 
Hirata2007, Okumura2009a, Okumura2009b, Joachimi2011, Mandelbaum2011, 
Singh2015, Singh2016, Uitert2017}.

Therefore, a major task in weak lensing cosmology is to remove/alleviate IA.  
There have been various proposals.  
From the data side,  the II term can be eliminated by removing close galaxy pairs with the
aid of photo-$z$ information, or disregarding auto correlation within
the same photo-$z$ bin \citep{King2002, King2003, Heymans2003, Takada2004, King2005}. 
However, this results in significant loss of information. 
Furthermore, it does not eliminate the GI term.  
From the IA theory side, one may adopt specific models of IA, 
and fit IA model parameters simultaneously with cosmological parameters.  
This approach has been applied in CFHTLens, KiDS, and
DES \citep[e.g.][]{Kirk2010, Heymans2013, Abbott2016, Hildebrandt2017, Joudaki2017, Troxel2017}. 
The main problem is the induced dependence on  IA modeling. 
The nulling technique avoids such model dependence, by introducing a
redshift-dependent weighting scheme to suppress the IA contribution \citep{Joachimi2008, Joachimi2009}. 
However, by design, the same weighting results in significant
loss of weak lensing information, in particular, its redshift dependence. 

\citet{Zhang2010a, Zhang2010b} proposed two self-calibration
techniques of both the GI and II contamination. 
The key in these self-calibration techniques is the discovered scaling relations independent of IA modeling. 
These scaling relations connect statistics of IA spatial clustering
with that of the galaxy number density field and the matter density field (gravitational lensing). 
Combining all observables (galaxy shapes and galaxy number density)
available in the same weak lensing survey, 
these scaling relations allow for unique determination of GI, II, and the lensing power spectrum.  
These self-calibration techniques have been extended to three-point
statistics \citep{Troxel2012a, Troxel2012b, Troxel2012c,Troxel2015}. 
Recently \citet{Yao2017} showed that the self-calibration technique
can indeed render the otherwise significant IA contamination insignificant, 
without sacrificing the lensing signal and cosmological information.  

The scaling relations \citep{Zhang2010a, Zhang2010b, Troxel2012a, Troxel2012b, Troxel2012c, Troxel2015} 
have only been verified with several analytical IA models. 
The next step is to verify them in more realistic situations. 
It is therefore the main goal of the current paper, to test the three scaling relations 
proposed in \citet[][, hereafter Z10]{Zhang2010b} with a high resolution
$\Nbody$ simulation of $3072^3$ simulation particles. 
Observationally, the IA contamination may only be non-negligible for
early-type galaxies. The observed  GI and II  can both be well
explained by the spatially correlated halo ellipticities in $\Nbody$
simulations, together with misalignment between ellipticities of galaxies and
halos \citep{Okumura2009a, Okumura2009b}.  
One can prove that the existence of galaxy misalignment 
does not affect the above scaling relations.  
We have randomly rotated halos to present the  misalignment of galaixes, 
and it only suppresses the IA amplitude, which agrees with previous works \citep[e.g.][]{Joachimi2013}.
Therefore, we carry out direct tests of scaling relations for halo ellipticities 
and their validity automatically applies to realistic early-type galaxies. 
With the verification of these scaling relations, 
the self-calibration technique in Z10 is  now complete from the theory side. 

This paper is organized as follows. \S 2 describes the
self-calibration technique. \S 3 describes the simulation and data
analysis. \S 4 verifies the scaling relations proposed in Z10.  Three
new scaling relations are proposed and verified too.  \S 5
discusses and summarizes.

%%%%%%%%%%%%%%%%%%%%%%%%%%%%%%%%%%%%%%%%%%%%%%%%%%%%%%%%
%%%%%%%%%%%%%%%%%%%%%%%%%%%%%%%%%%%%%%%%%%%%%%%%%%%%%%%%
\section{ The self-calibration technique}
\label{sec:tech}
%%%%%%%%%%%%%%%%%%%%%%%%%%%%%%%%%%%%%%%%%%%%%%%%%%%
Here we briefly summarize the self-calibration technique proposed in Z10. 
Conventional lensing tomography usually adopts coarse redshift bins of width $\sim 0.2$, 
comparable to photo-$z$ error of individual galaxies. 
Such a coarse bin size is sufficient for extracting the cosmological information in weak lensing, 
since the lensing kernel only varies slowly with redshift. 
However, it disregards information valuable for calibrating IA. 
Due to the large number of source galaxies, even much finer photo-$z$ bins
(e.g.  of width $\sim 0.01$) may have millions of galaxies, 
and therefore can have sufficient S/N of diagnosing IA. 
The IA self-calibration utilizes such information.
First, we split galaxies into redshift bins of narrow width
$\mathcal{O}(0.01)$ and work on various
two-point cross-correlations between photo-$z$ bins. 
Unless otherwise specified, we focus on the E-mode of galaxy ellipticities in cosmic shear surveys. 
The observed E-mode galaxy ellipticity is 
\ba
\gamma^{\rm O}_E=\gamma+I\ . \no
\ea
Here $\gamma$ is the cosmic shear and $I$ is the E-mode galaxy IA. 
The spatially uncorrelated (random) component of the galaxy shapes does not 
contribute in the nonzero lag correlation function, 
and the shape noise can be subtracted in the observed power spectrum. 
Therefore, for brevity we ignore this component.  
Together with the other observable, namely the galaxy surface number density, 
we can  form three two-point auto/cross-correlations. 
The corresponding power spectra between the $i$th and the $j$th redshift bins are 
$C^{(1)}_{ij}$ (the power spectrum between the observed(lensed) galaxy ellipticity), 
$C^{(2)}_{ij}$ (the galaxy ellipticity-galaxy surface overdensity cross-power spectrum), 
and  $C^{(3)}_{ij}$ (the galaxy surface overdensity power spectrum). 
Notice that all three power
spectra are symmetric with respect to the $ij$ pair ($C^{(1,2,3)}_{ij}=C^{(1,2,3)}_{ji}$).  
\ba
\label{eqn:correlations} 
C_{ij}^{(1)}(\ell) &=& C_{ij}^{\rm{GG}}(\ell) + C_{ij}^{\rm{II}}(\ell) +
C_{ij}^{\rm{GI}}(\ell) + C_{ij}^{\rm{IG}}(\ell) \ , \no\\
C_{ij}^{(2)}(\ell) &=& C_{ij}^{\rm{Gg}}(\ell) + C_{ij}^{\rm{gG}}(\ell)
+ 2C_{ij}^{\rm{Ig}}(\ell) \ ,\no\\
C_{ij}^{(3)}(\ell)&=& C_{ij}^{\rm{gg}}(\ell) \ .
\ea
These measurable quantities are related to the underlying power spectra $C^{\alpha\beta}_{ij}$. 
Here, $\alpha,\beta= G,  I,  g$ denote the gravitational shear, 
the IA, and the galaxy surface overdensity respectively. 
$C^{\alpha\beta}_{ij}$ is the cross-power spectrum between the property $\alpha$ of the $i$th
redshift bin and the property $\beta$ of the $j$th redshift bin. 
When $\alpha=\beta$, $C^{\alpha\beta}_{ji}=C^{\alpha\beta}_{ij}$. 
However, when $\alpha=G$ and $\beta=I,g$, $C^{\alpha\beta}_{ji}\neq C^{\alpha\beta}_{ij}$. 

%%###########################################%%
\begin{figure}
\centering
\includegraphics[width=0.48\textwidth]{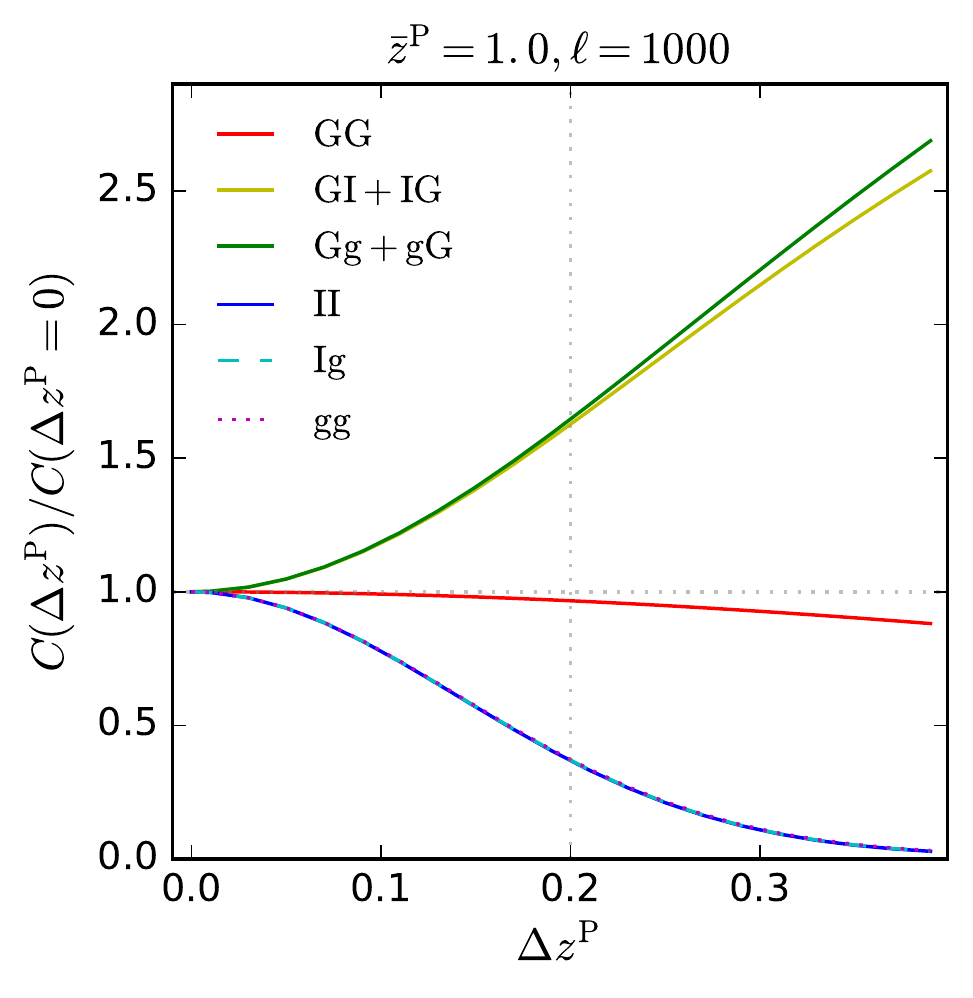}
\caption{Classifications of various power spectra in the $\Delta z^{\rm P}$ space from the $\Nbody$ simulation data. 
The cross-power spectra between galaxy IA/number overdensity and gravitational 
lensing increase with increasing $\Delta z^{\rm P}$. 
In contrast, the power spectra between galaxy IA and number overdensity decreases 
with increasing $\Delta z^{\rm P}$. 
The relative dependences are almost identical for II, Ig, and gg, 
and are therefore indistinguishable in the plot.  
The lensing power spectrum $C^{\rm GG}$ only weakly depends on $\Delta z^{\rm P}$. 
This significant difference in the $\Delta z^{\rm P}$ dependence, 
along with the three scaling relations shown in Figures \ref{fig:S1}-\ref{fig:S3}, 
are key ingredients of the proposed IA self-calibration technique. 
}
\label{fig:DZpDependence}
\end{figure}
%%###########################################%%

In the measurement $C^{(1)}_{ij}$, the lensing power spectrum $C^{\rm GG}_{ij}$  
is contaminated by both the II term and the GI term from the galaxy IA. 
Adding new measurements ($C^{(2,3)}_{ij}$) on one hand provides extra constraints on IA, 
but on the other hand  brings extra unknown quantities.  
Z10 found several scaling relations between these unknown quantities. 
They  significantly reduce the degrees of
freedom in Equation (\ref{eqn:correlations}), and make $C^{\rm GG}$ solvable. 
These scaling relations are generic, arising from
the very basic fact that both IA and the galaxy number density are
intrinsic 3D fields, while cosmic shear is the projection of a 3D field
(matter density). The self-calibration based on them is therefore
independent of IA modeling.

We denote the average redshift of galaxies in the $i$th redshift bin as $z_i^{\rm P}$. 
Here the superscript ``P'' denotes the photometric redshift.  
The above power spectra then depend on both $z_i^{\rm P}$ and $z_j^{\rm P}$. 
Such dependences can be re-expressed as the dependences on the mean
redshift $\bar{z}^{\rm P}\equiv (z^{\rm P}_i+z^{\rm P}_j)/2$, and the
redshift separation $\Delta z^{\rm P}\equiv z^{\rm P}_j-z^{\rm P}_i$. 
With respect to these new arguments ($\Delta z^{\rm P}$,
$\bar{z}^{\rm P}$ and $\ell$),  the scaling relations found in Z10 are 
\ba
\label{eqn:S}
{\rm{S1}}:& C^{\rm {II}}(\Delta z^{\rm P}|\ell,\bar{z}^{\rm P}) \simeq
A_{\rm {II}}(\ell,\bar{z}^{\rm P})C^{\rm {gg}}(\Delta z^{\rm
  P}|\ell,\bar{z}^{\rm P}) \ , \no\\
{\rm{S2}}:& C^{\rm {Ig}}(\Delta z^{\rm P}|\ell,\bar{z}^{\rm P})\simeq A_{\rm {Ig}}(\ell,\bar{z}^{\rm P})C^{\rm {gg}}(\Delta
z^{\rm P}|\ell,\bar{z}^{\rm P}) \ ,\no\\
{\rm{S3}}:& C^{\rm{GI}}(\Delta z^{\rm P}|\ell,\bar{z}^{\rm
  P})+C^{\rm{IG}}(\Delta z^{\rm P}|\ell,\bar{z}^{\rm P}) \simeq A_{\rm {GI}}(\ell,\bar{z}^{\rm P})\no\\
& \times \left[C^{\rm {Gg}}(\Delta z^{\rm
    P}|\ell,\bar{z}^{\rm P})+C^{\rm {gG}}(\Delta z^{\rm
    P}|\ell,\bar{z}^{\rm P})\right]  \ .
\ea
The prefactors $A_{\rm{II}}$, $A_{\rm{Ig}}$, and
$A_{\rm{GI}}$ encode information of IA, but are hard to calculate from
the first principle.  What the scaling relations emphasize is that, $A_{\rm{II}}$, $A_{\rm{Ig}}$ and
$A_{\rm{GI}}$ do not depend on $\Delta z^{\rm P}$. 
In other words, the two sets of power spectra in the S1$\sim$S3 scaling relations  have identical
$\Delta z^{\rm P}$ dependence. 
Namely, the ratios (e.g. $C^{\rm II}/C^{\rm gg}$ for fixed $\ell$ and $\bar{z}^{\rm P}$) 
should be $\Delta z^{\rm P}$ independent. Z10 predicts that this should be valid at $\Delta z^{\rm P}\la 0.2$.

%%###########################################%%
\begin{figure}
\includegraphics[width=0.48\textwidth]{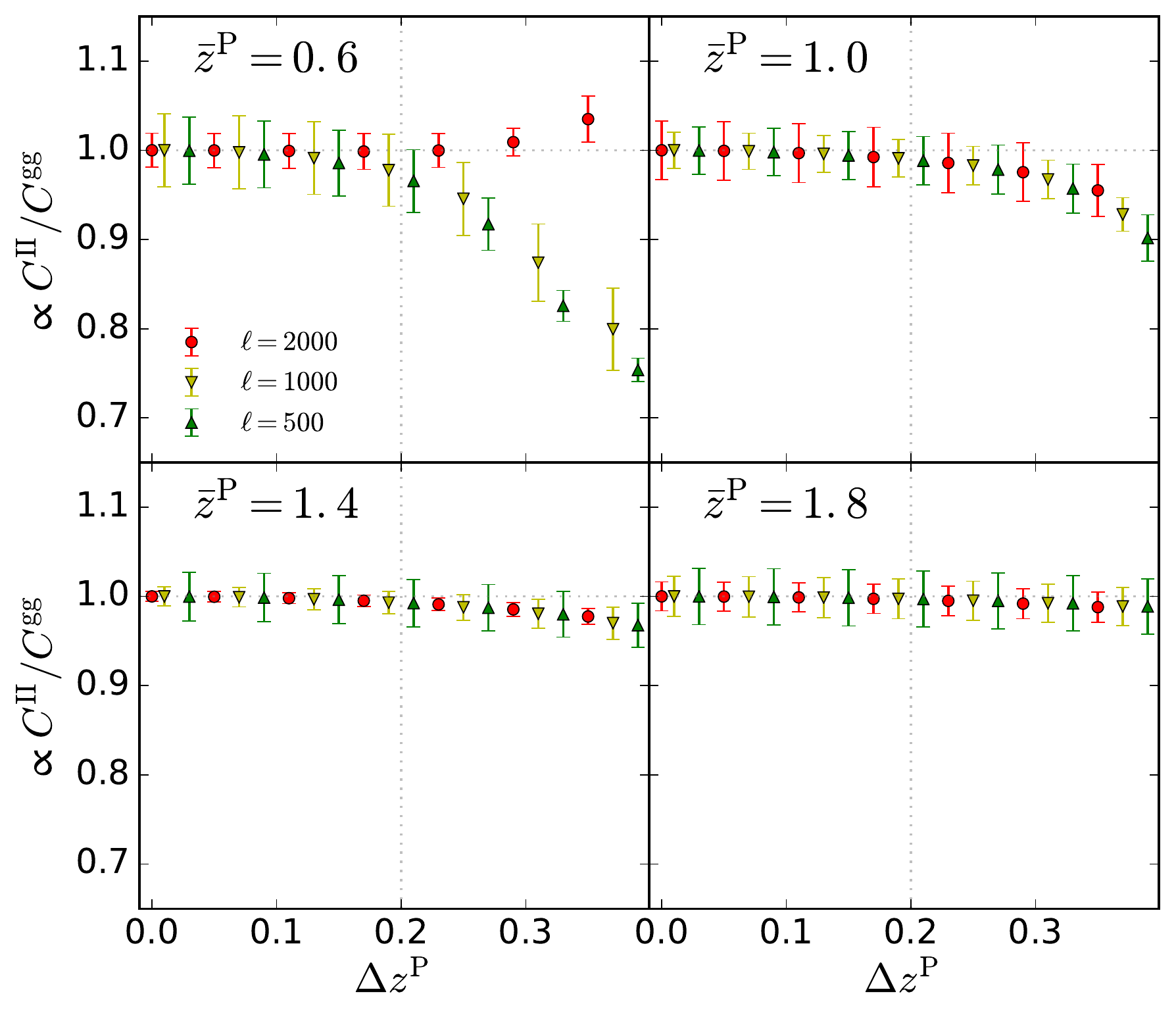}
\caption{Verification of the scaling relation S1. 
S1 states that the ratio $C^{\rm{II}}(\Delta z^{\rm P})/C^{\rm{gg}}(\Delta z^{\rm P})$ 
should be independent of $\Delta z^{\rm P}$. 
Since the absolute amplitude is irrelevant, 
we scale the first data points (at $\Delta z^{\rm P}=0$) to unity.  
S1 holds to an accuracy of $\mathcal{O}(1)\%$ at $\Delta z^{\rm P}\leq 0.2$ and 
$\bar{z}^{\rm P}=0.6$. 
Its accuracy further improves toward higher $\bar{z}^{\rm P}$. 
}
\label{fig:S1}
\end{figure}
%%###########################################%%
Fig. \ref{fig:DZpDependence} shows the $\Delta z^{\rm P}$ dependences
of the above power spectra measured from our simulation detailed later, 
for $\ell=1000$ and $\bar{z}^{\rm P}=1.0$. 
Although this figure indeed shows the validity of the above scaling relations, 
the major purpose is to demonstrate how the self-calibration works. 
The $\Delta z^{\rm P}$ dependences can be naturally classified into
three categories, insensitive to  details of IA. 
\bi
\item 
The lensing power spectrum only weakly depends on $\Delta z^{\rm P}$, 
since the lensing kernel varies slowly with redshift. 
Based on this slow variation,  Z10 also derived a generic scaling relation on the lensing power spectrum, 
\ba
\label{eqn:GG}
\frac{C^{\rm {GG}}(\Delta z^{\rm P}|\ell,\bar{z}^{\rm P})}{C^{\rm {GG}}(0|\ell,\bar{z}^{\rm P})} \simeq 1-A_{\rm
  GG}(\ell, \bar{z}^{\rm P})(\Delta z^{\rm P})^2\ .
\ea 
\item 
$C^{\rm II}$, $|C^{\rm Ig}|$, and $C^{\rm gg}$ decrease quickly with increasing $\Delta z^{\rm P}$. 
This is the natural consequence of short correlation length of
the underlying 3D fields of IA and galaxy number density. 
\item 
In contrast, both $|C^{\rm GI}+C^{\rm IG}|$ and $C^{\rm Gg}+C^{\rm gG}$ 
increase quickly with increasing $\Delta z^{\rm P}$. 
This is the natural consequence of
higher lensing efficiency for larger source-lens separation.
\ei
 
 Observationally, we can find redshift
bin pairs of identical $\bar{z}^{\rm P}$ but different $\Delta z^{\rm P}$.  
This allows us to separate the three categories of components using their different
$\Delta z^{\rm P}$ dependences. 
Mathematically, with measurements of $C^{(1,2,3)}$ at four or more $\Delta z^{\rm P}$, 
we are able to solve for all unknowns in Equations (\ref{eqn:correlations})-(\ref{eqn:GG}). 
We then obtain $C^{\rm GG}$, free of IA contamination, and independent of IA modeling. 
What's more, we can get all the prefactors $A_{\alpha \beta}$ in Equation (2)-(3) from all the observed quantities. 
So on the validation of these scaling relations we do not care the value of these prefactors 
$A_{\alpha \beta}$ which indeed depend on the detailed IA physics.

Equation (\ref{eqn:GG}) has been verified unambiguously in Z10, 
since we already have a sufficiently accurate understanding of weak lensing statistics. 
The same paper also demonstrated the robustness of the S1$\sim$S3 scaling relations,
but only for specific analytical/semianalytical IA models. 
The major remaining question is whether S1$\sim$S3 hold in more realistic situations. 
Therefore, we test these scaling relations with IA in numerical simulations. 

%%###########################################%%
\begin{figure}
\centering
\includegraphics[width=0.48\textwidth]{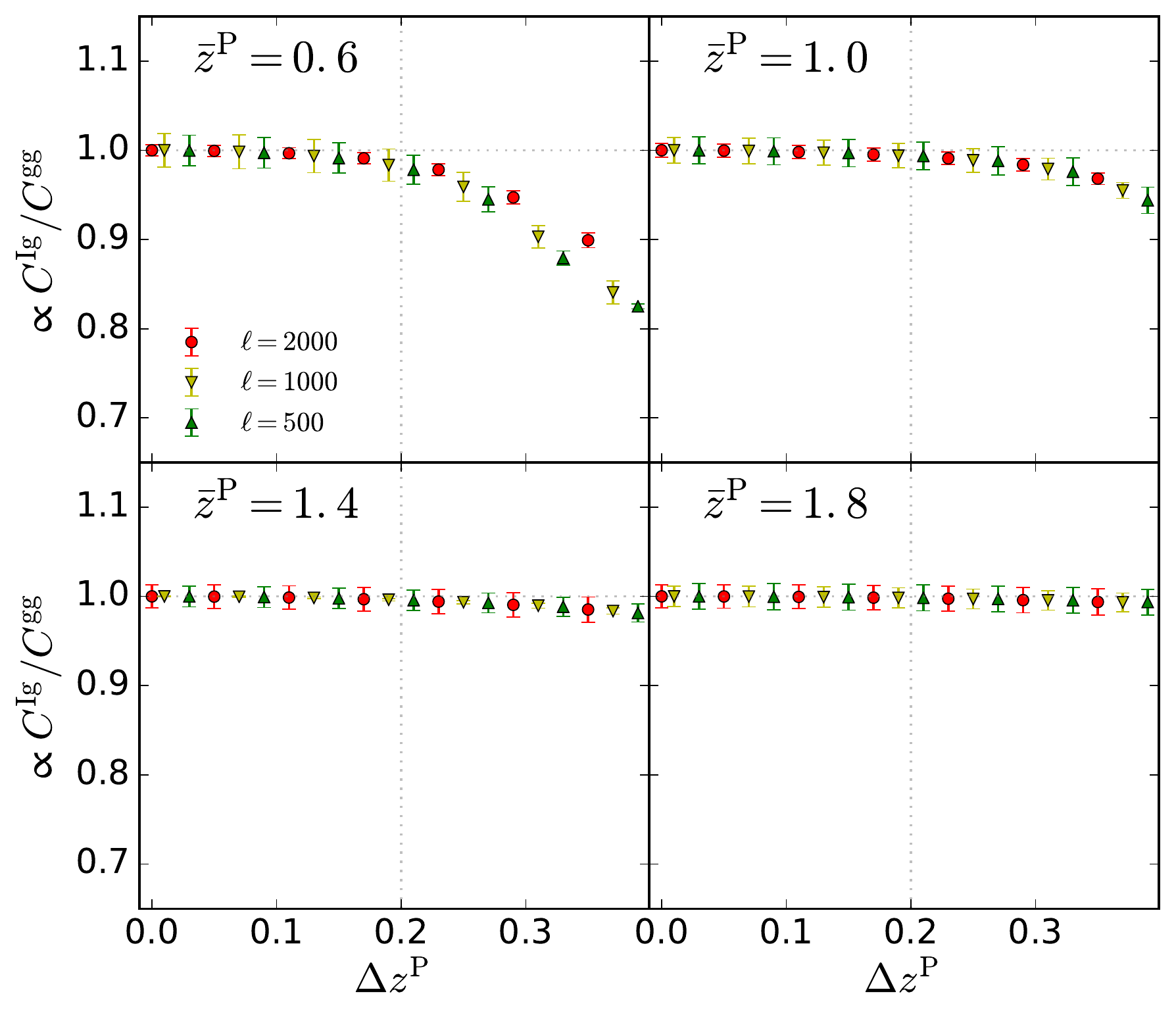}
\caption{Verification of the scaling relation S2. 
It is similar to Figure \ref{fig:S1}, but for the ratios $C^{\rm{Ig}}/C^{\rm{gg}}$. 
}
\label{fig:S2}
\end{figure}
%%###########################################%%
%%%%%%%%%%%%%%%%%%%%%%%%%%%%%%%%%%%%%%%%%%%%%%%%%%%%%%%%
%%%%%%%%%%%%%%%%%%%%%%%%%%%%%%%%%%%%%%%%%%%%%%%%%%%%%%%%
\section{The $\Nbody$ Simulation and Data Analysis}
\label{sec:data}
%%%%%%%%%%%%%%%%%%%%%%%%%%%%%%%%%%%%%%%%%%%%%%%%%%%%%%%%
The $\Nbody$ simulation we analyze has $3072^{3}$ dark matter
particles in  a  $(600 \mpch)^3$ cosmic volume (hereafter J6610). 
It was run with a particle-particle-particle-mesh (${\rm P}^{3} {\rm M}$) code \citep{Jing2007,  Jing2018}. 
It adopts a flat $\LCDM$ cosmology with $\omegam$ = $1-\omegal$ = 0.268, $\omegab= 0.045$,  
$h\equiv H_{0}/ 100 / \kmsmpc= 0.71$,  $\sigma_{8}=0.83$, and $n_s=0.968$.  
The dark matter halos are first identified using the friends-of-friends algorithm 
with a linking length $20\%$ of the mean particle separation.  
All unbound particles are excluded in the final halo catalog. 
Our work only uses halos with at least 20 simulation particles. 
We restrict to the IA of early-type galaxies. 
Their IA is expected to arise from the ellipticities of host halos, 
up to a misalignment angle. 
The misalignment reduces the IA amplitude. 
However, it does not change the scaling relations S1$\sim$S3, 
as long as misalignments of galaxies in different halos are spatially uncorrelated. 
For this reason, we only need to test the scaling relations for halos and the results 
obtained automatically apply to early-type galaxies. 

The halo ellipticities are defined (e.g.  for projection onto the $x$-$y$ plane), 
\be
\epsilon_1 = \frac{I_{xx} - I_{yy}}{I_{xx} + I_{yy}} \ ,  \   \
\epsilon_2 = \frac{2I_{xy}}{I_{xx} + I_{yy}} \ .
\label{eq:ellipticity}
\ee
$I_{\alpha\beta}$ is the inertia tensor, 
\be
I_{\alpha \beta} = \frac{\Sigma_i^{N} w_{i} m_{i} (\alpha_i - \bar{\alpha}) (\beta_i - \bar{\beta})}{\Sigma_{i}^{N} w_{i} m_{i}}.
\label{eq:IT}
\ee
Here $\alpha_{i}$ and $\beta_{i}$ are the coordinates of the $i$th halo particle in the simulation box,
and $\bar{\alpha}$ and $\bar{\beta}$ are the coordinates of the halo center. 
$m_{i}$ is the mass of the $i$th particle and $w_{i}$ is its weighting. 
We adopt $w_{i} = 1/r_{i}^{2}$, where $r_{i}$ is the distance of this
particle to the halo center.
$I_{\alpha\beta}$ defined with such weighting is called the reduced inertia tensor. 
We then have the 3D distribution of ellipticities. 
With respect to the same line of sight, we can perform
the E-B separation and obtain the 3D distribution of E-mode intrinsic ellipticity ($I$). 

%%###########################################%%
\begin{figure}
\centering
\includegraphics[width=0.48\textwidth]{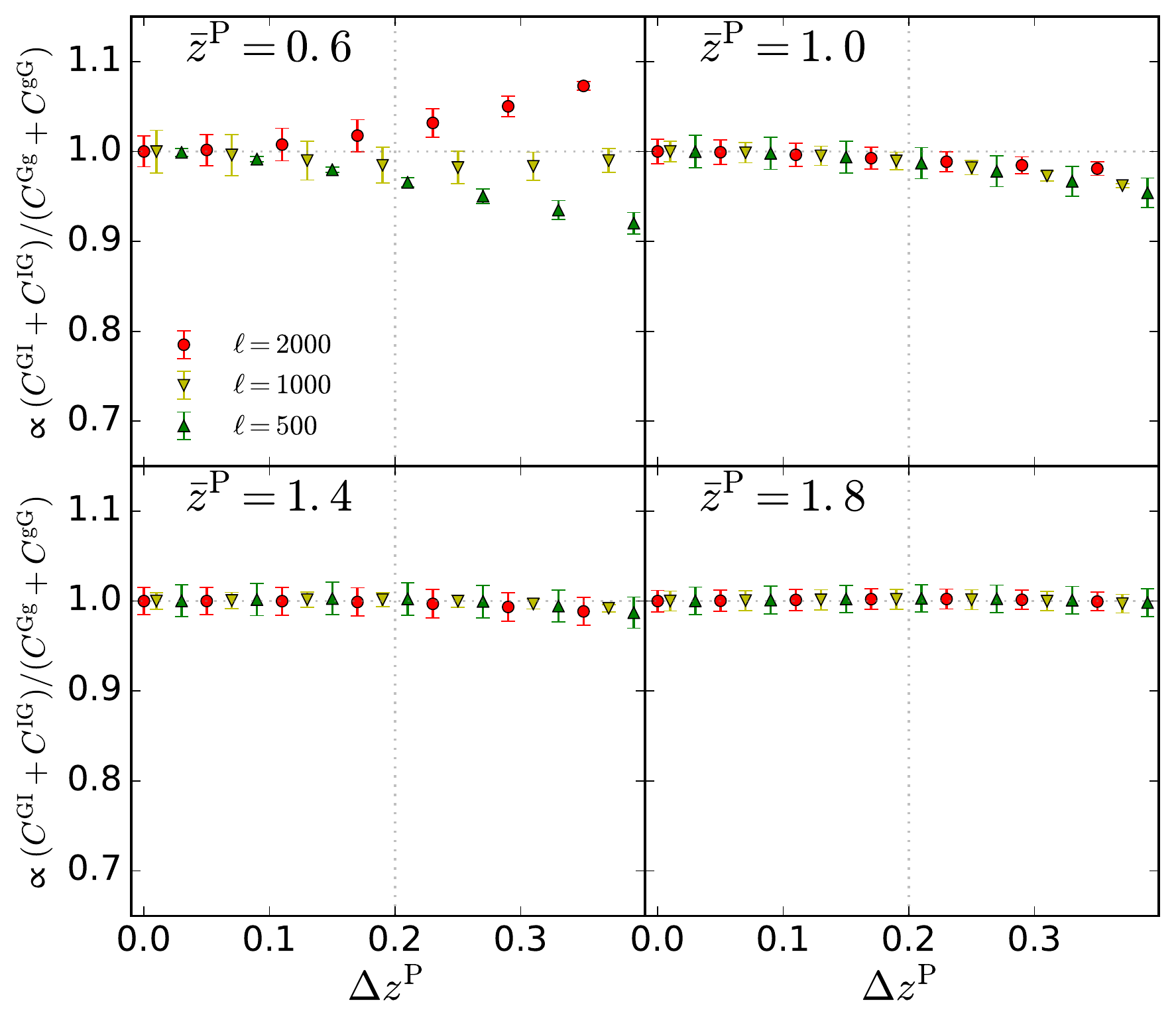}
\caption{Verification of the scaling relation S3. 
It is similar to Figure \ref{fig:S1}, but for the ratios 
$(C^{\rm{GI}}+C^{\rm{IG}})/(C^{\rm{Gg}}+C^{\rm{gG}})$.  
}
\label{fig:S3}
\end{figure}
%%###########################################%%
%%###########################################%%

There are two ways to calculate $C^{\alpha\beta}(\ell)$ in the previous section. 
One way is to first make maps of $\alpha=G, I, g$, and then measure their cross-power spectra. 
This is straightforward. 
However, the large photo-$z$ error causes large scatter
in the maps and we need to produce many of them to reduce the
statistical fluctuation in $C^{\alpha\beta}$.  
Since these maps are not independent of each other, 
it is then difficult to quantify the statistical error in $C^{\alpha\beta}$.
Another way is to first measure the corresponding 3D power spectra
$P^{\alpha\beta}(k,z)$ from the $I$, $\delta_m$, and $\delta_g$ fields
and then apply the Limber integral to obtain $C^{\alpha\beta}$. 
\ba
\label{eqn:Cab}
\frac{\ell^2}{2\pi}C^{\alpha \beta}(\Delta z^{\rm P}|\ell, \bar{z}^{\rm P}) = \frac{\pi}{\ell} \int_{0}^{\infty} 
\Delta^{2}_{\alpha \beta}\left(k=\frac{\ell}{\chi(z)}, z\right) \no\\
\times W_{\alpha \beta}(z, \Delta z^{\rm P}, \bar{z}^{\rm P})
\tilde{\chi}(z) \frac{H(z)}{H_{0}} {\rm d}z \ . 
\ea
Here, $\Delta^{2}_{\alpha \beta}\equiv k^{3}P_{\alpha \beta}(k)/2\pi^{2}$ is
the corresponding (dimensionless) 3D power spectrum variance.
$W_{\alpha\beta}$ is the corresponding weighting function, whose
details are given in Z10. 
$\tilde{\chi}(z) \equiv \chi/(c/H_{0})$ is the comoving angular
diameter distance in units of the Hubble radius. 
$H(z)$ is the Hubble parameter at redshift $z$. 
Since the measurement of $P^{\alpha\beta}$ uses all two-point information 
in the whole simulation box, 
the obtained $C^{\alpha\beta}$ has minimal statistical fluctuations. 
Therefore, we will adopt this approach. For each line of sight ($x$, $y$, $z$), 
we have an independent $I$ field and its corresponding power spectra $P^{I\alpha}$ and $C^{I\alpha}$. 
Comparing between the three lines of sight, 
we obtain the statistical errors of $C^{I\alpha}$ and then quantify
the accuracy of scaling relations. But we caution that such errorbars
should not be used for cosmological constraints, since random shape noise is not included.

%%###########################################%%
\begin{figure}
\centering
\includegraphics[width=0.48\textwidth,height=0.43\textwidth]{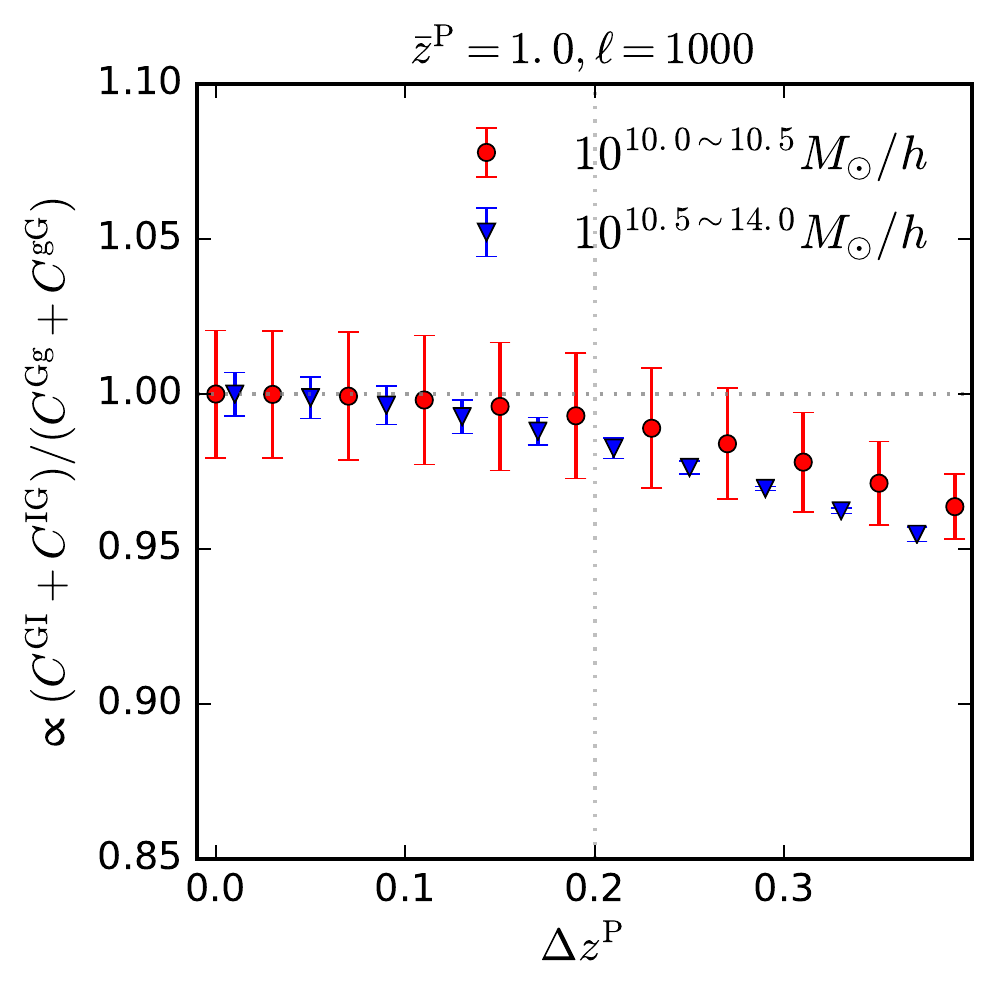}
\caption{Scaling relations hold for different halo masses. 
This figure is similar to Figure \ref{fig:S3}, but for two halo mass bins.
}
\label{fig:DiffMass}
\end{figure}
%%###########################################%%

%%%%%%%%%%%%%%%%%%%%%%%%%%%%%%%%%%%%%%%%%%%%%%%%%%%%%%%%
%%%%%%%%%%%%%%%%%%%%%%%%%%%%%%%%%%%%%%%%%%%%%%%%%%%%%%%%
\section{Verifications of the scaling relations}
\label{sec:verification}
%%%%%%%%%%%%%%%%%%%%%%%%%%%%%%%%%%%%%%%%%%%%%%%%%%%%%%%%
We test the scaling relations (S1$\sim$S3) over a variety of redshifts and angular scales. 
For brevity, we only show the results at $\ell = 500, 1000, 2000$ of interest in weak lensing cosmology. 
We also adopt $\bar{z}^{\rm P} = 0.6, 1.0, 1.4, 1.8$, which cover a large
range of accessible source redshifts for most weak lensing surveys. 

Fig. \ref{fig:S1} plots the ratio  $C^{\rm II}(\Delta z^{\rm P})/C^{\rm gg}(\Delta z^{\rm P})$ 
as a function of $\Delta z^{\rm P}$ for fixed $\ell$ and $\bar{z}^{\rm P}$ with values listed above. 
If the scaling relation S1 holds, the ratios
should be horizontal lines (independent of $\Delta z^{\rm P}$). 
Fig. \ref{fig:S1} shows that this is indeed the case. 
The overall accuracy reaches $\mathcal{O}(1)\%$ at $\Delta z^{\rm P}\la 0.2$. 
The accuracy is better for larger $\ell$ and higher $\bar{z}^{\rm P}$,
as predicted in Z10. For example, $1\%$ accuracy remains for $\Delta z^{\rm P}\leq 0.3$,  
for $\ell=2000$ and $0.6\leq \bar{z}^{\rm P}\leq 1.8$,  and for $500\leq \ell\leq 2000$ and $\bar{z}^{\rm P}=1.4/1.8$ .

Fig. \ref{fig:S2} shows the ratios $C^{\rm Ig}/C^{\rm gg}$, 
and verifies the scaling relation S2. 
Fig. \ref{fig:S3} shows the ratios $(C^{\rm GI}+C^{\rm IG})/(C^{\rm Gg}+C^{\rm gG})$ 
and verifies the scaling relation S3.  
The accuracies of S2 and S3 also reach $\mathcal{O}(1)\%$,
comparable to that of S1. Therefore, we verify the scaling relations
for halo ellipticities. The ellipticities of early-type galaxies can
be well described by the halo ellipticities and a spatially
uncorrelated misalignment angle between halos and galaxies
\citep{Okumura2009a, Okumura2009b}. Such misalignment does not affect
the S1$\sim$S3 scaling relations.   We then conclude that  the
predicted scaling relation S1$\sim$S3 are accurate 
to the $\mathcal{O}(1)\%$ level when $\Delta z^{\rm P} \la 0.2$, 
for both halo and galaxy ellipticities. 
Such a level of accuracy is sufficient for accurate removal of IA in cosmic shear power spectrum measurement. 

%%###########################################%%
\begin{figure}
\centering
\includegraphics[width=0.48\textwidth,height=0.43\textwidth]{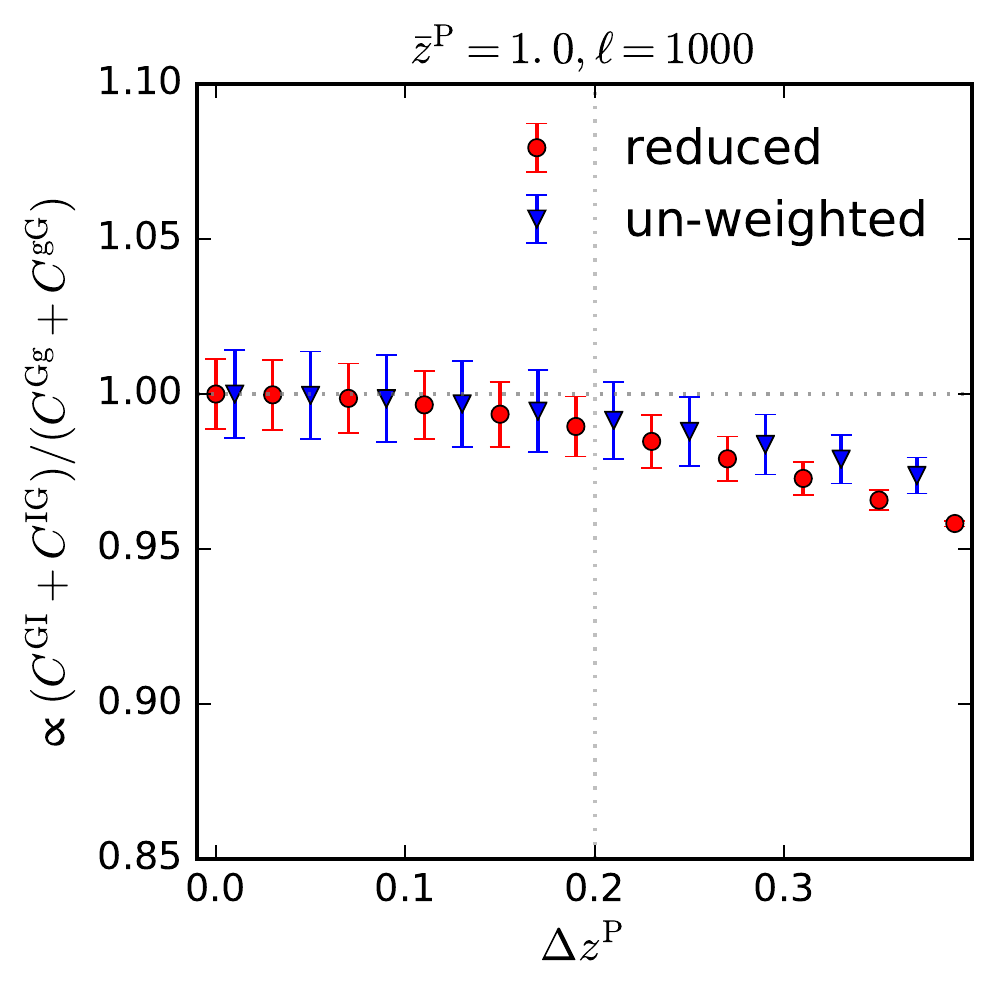}
\caption{Scaling relations hold for different definitions of halo ellipticities. 
This figure is similar to Figure \ref{fig:S3}, but for comparing two
different ways of calculating the inertia tensor (reduced and
unweighted) and the halo ellipticities.}
\label{fig:DiffRweight}
\end{figure}
%%###########################################%%

%%%%%%%%%%%%%%%%%%%%%%%%%%%%%%%%%%%%%%%%%%%%%%%%%%%%%%%%
\subsection{The universality of the scaling relations}
\label{subsec:massweight}
%%%%%%%%%%%%%%%%%%%%%%%%%%%%%%%%%%%%%%%%%%%%%%%%%%%%%%%%
The scaling relations S1$\sim$S3 are universal in several ways. 
They hold for halos of different mass. 
We split halos into two mass bins, $10^{10}<M/(M_{\odot}/h)<10^{10.5}$ and
$10^{10.5}<M/(M_{\odot}/h)<10^{14}$ and redo the
tests.\footnote{\citet{Jing2002} pointed out for halos with too few particles,  
the ellipticity correlation will be underestimated, 
and the underestimation amounts to a factor of two for halos of 20 simulation particles. 
This underestimation can be described by an extra misalignment. 
Therefore, it does not change the scaling relations. 
The low-mass halo sample verifies this argument. }
For brevity, we only show the test against S3 at $\bar{z}^{\rm P}=1.0$ and $\ell=1000$ (Fig. \ref{fig:DiffMass}). 
In addition, scaling relations S1, S2, and other $\bar{z}^{{\rm P}}$, $\ell$ have similar conclusions.
Both halo samples obey the scaling relation S3 to $\mathcal{O}(1)\%$ 
accuracy at $\Delta z^{\rm P}\leq 0.2$ ($\mathcal{O}(5)\%$ accuracy at $\Delta z^{\rm P}\leq 0.4$). 

They hold for different weighting $w_i$ in the definition of the
inertial tensor and the ellipticities (Eq. \ref{eq:IT}).  
Previous works \citep[e.g.][]{Pereira2008, Joachimi2013, Hilbert2017} 
showed that ellipticities can differ significantly in the inner and
outer regions of halos/galaxies and therefore depend significantly on the weighting $w_i$. 
However, this does not affect the scaling relations S1$\sim$S3. 
Figure \ref{fig:DiffRweight} shows the tests of S3 adopting $w_{i} = 1$ 
and compare it with the case of $w_i=1/r_i^2$ adopted previously. 
Despite the strong dependence of IA on the weighting, the scaling relations still hold. 
This further supports the argument in Z10 that scaling relations S1$\sim$S3 
are generic  and do not rely on the details of IA.

They are also insensitive to details of photo-z error distribution. 
All the power spectra $C^{\alpha\beta}$ are affected by the photo-$z$ errors.  
But since the photo-$z$ errors affect both the left- and right-hand sides in basically the same way, 
the scaling relations S1$\sim$S3 are insensitive to the photo-$z$ errors. 
The above tests adopt a double Gaussians PDF for photo-z errors, detailed in Z10. 
We have checked that, for a single Gaussian or non-Gaussian PDF,  the scaling relations also hold. 

  %%###########################################%%
\begin{figure}
\centering
\includegraphics[width=0.48\textwidth]{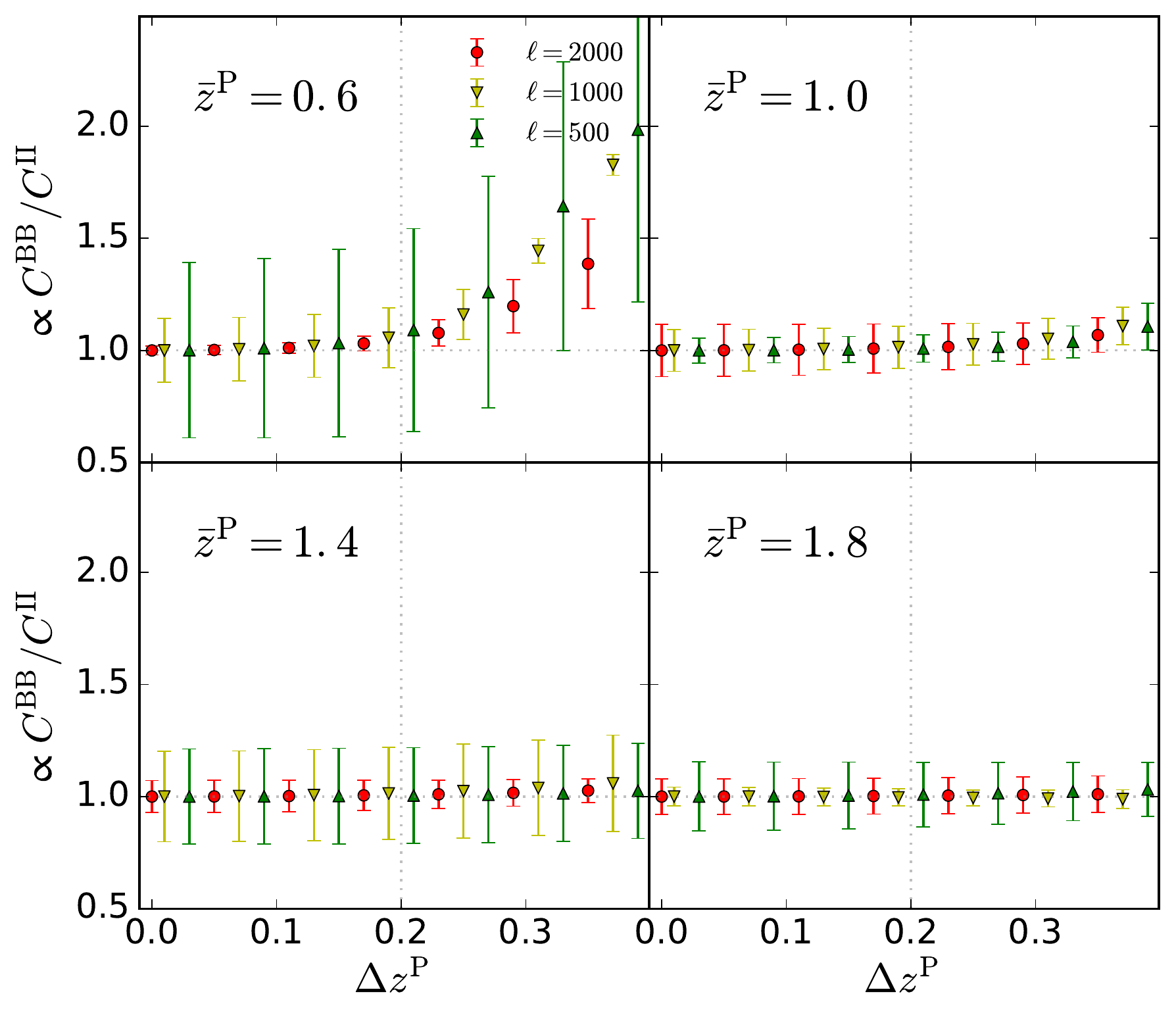}
\caption{Verification of the B-mode scaling relation proposed in Eq. \ref{eqn:S4} .  It is similar to
  Fig. \ref{fig:S1}, but for the B-E relation. These
  new scaling relations are useful when studying the IA B-mode as a
  tracer of the large-scale structure and when calibrating IA in the weak lensing
  measurement.  Notice that errorbars at different $\Delta z^{\rm P}$
  are correlated due to the common 3D power spectra shared in the
  Limber integral (Equation (\ref{eqn:Cab})). 
}
\label{fig:S4}
\end{figure}
%%###########################################%%
%%%%%%%%%%%%%%%%%%%%%%%%%%%%%%%%%%%%%%%%%%%%%%%%%%%
%%%%%%%%%%%%%%%%%%%%%%%%%%%%%%%%%%%%%%%%%%%%%%%%%%%%%%%%
\subsection{Scaling relations of B-mode IA}
\label{subsec:Bmode}
%%%%%%%%%%%%%%%%%%%%%%%%%%%%%%%%%%%%%%%%%%%%%%%%%%%%%%%%
Unlike cosmic shear, which has negligible B-mode, B-mode of IA is
usually non-negligible (e.g. \citet{Crittenden2001, Crittenden2002, Hirata2004b, Heymans2006b}). 
By its symmetry property, B-mode is only correlated with itself and the only nonvanishing two-point
statistics is its auto power spectrum $C^{\rm BB}$. 
Based on the same argument in Z10, we expect the following three scaling relations, 
\ba
\label{eqn:S4}
{\rm{S4}}: C^{\rm {BB}}(\Delta
z^{\rm P}|\ell,\bar{z}^{\rm P}) &\simeq& A_{\rm
  {BE}}(\ell,\bar{z}^{\rm P}) C^{\rm {II}}(\Delta z^{\rm
  P}|\ell,\bar{z}^{\rm P}) \no \\
&\simeq& A_{\rm
  {BIg}}(\ell,\bar{z}^{\rm P}) C^{\rm {Ig}}(\Delta z^{\rm
  P}|\ell,\bar{z}^{\rm P}) \no \\
&\simeq& A_{\rm
  {Bg}}(\ell,\bar{z}^{\rm P}) C^{\rm {gg}}(\Delta z^{\rm
  P}|\ell,\bar{z}^{\rm P})\ .
\ea
The same simulation also verifies the above three new scaling
relations, and for brevity we only use one of the tests (Fig. \ref{fig:S4}).
These scaling relations are useful in both separating shape measurement errors  in
B-mode  (e.g. RCSLenS \citep{Hildebrandt2016}, KiDS-450 \citep{Hildebrandt2017}) 
and calibrating the IA E-mode .

%%%%%%%%%%%%%%%%%%%%%%%%%%%%%%%%%%%%%%%%%%%%%%%%%%%%%%%%
%%%%%%%%%%%%%%%%%%%%%%%%%%%%%%%%%%%%%%%%%%%%%%%%%%%%%%%%
\section{Discussion and Summary}
\label{sec:dis_sum}
%%%%%%%%%%%%%%%%%%%%%%%%%%%%%%%%%%%%%%%%%%%%%%%%%%%%%%%%
We carry out comprehensive tests of  the three scaling relations
useful for the IA self-calibration proposed in Z10. 
We conclude that these scaling relations are valid for all investigated angular scales $\ell$, 
mean-source redshifts $\bar{z}^{\rm P}$, source redshift separations $\Delta z^{\rm P}$, 
halo masses, weightings in the halo ellipticity definition, and photo-$z$ error distribution.  
The tests are for halo ellipticities. 
But since misalignment between ellipticities of early-type galaxies and
host halos does not alter these scaling relations,  the same conclusion
applies to galaxy ellipticities. This makes the proposed IA self-calibration complete on the theory side, 
and makes it ready for application to real data analysis of galaxy ellipticities and cosmic shear. 

In the original proposal of Z10, only scaling relations related to E-mode ellipticity are discussed. 
The same argument naturally  leads to scaling relations of B-mode ellipticity. 
We list these scaling relations and verify them  in \S \ref{subsec:Bmode}. 

There are further complexities to take care. For example, there are different estimators of measuring the ellipticity correlations. 
We adopt the pixel-based estimator. 
We also need to check the standard estimator \citep[e.g.][]{Munshi2008, Schmidt2009, Heymans2012}. 
Nevertheless, since the derivation of the scaling relations is not restricted to a specific estimator, 
we expect that these scaling relations should hold as well.  
Furthermore, the tests done for halos in $\Nbody$ simulations should be 
extended to galaxies in hydrodynamical simulations. 
This will allow us to more robustly check the dependence on galaxy types.

%%%%%%%%%%%%%%%%%%%%%%%%%%%%%%%%%%%%%%%%%%%%%%%%%%%
\section*{acknowledgements}
%%%%%%%%%%%%%%%%%%%%%%%%%%%%%%%%%%%%%%%%%%%%%%%%%%%
We thank Zheng Zheng, Jun Zhang, Huanyuan Shan, Le Zhang, Ji Yao, Xi Kang, and Peng Wang for useful discussions. 
This work was supported by the National Key Basic Research and Development Program of China (No. 2018YFA0404504), 
the National Science Foundation of China (11773048, 11403071, 11621303, 11433001, 11653003, 11533006, 11222325, 11033006 and 11320101002), 
the National Basic Research Program of China (2015CB857001, 2015CB857003), 
the Knowledge Innovation Program of CAS (KJCX2-EW-J01), 
Shanghai talent development funding (No. 2011069) and Shanghai Key Laboratory Grant (No.11DZ2260700). 
This work made use of the High Performance Computing Resource in the Core 
Facility for Advanced Research Computing at Shanghai Astronomical Observatory.

%%%%%%%%%%%%%%%%%%%%%%%%%%%%%%%%%%%%%%%%%%%%%%%%%%%
%%%%%%%%%%%%%%%%%%%%%%%%%%%%%%%%%%%%%%%%%%%%%%%%%%%
{}

%%%%%%%%%%%%%%%%%%%%%%%%%%%%%%%%%%%%%%%%%%%%%%%%%%%
%%%%%%%%%%%%%%%%%%%%%%%%%%%%%%%%%%%%%%%%%%%%%%%%%%%

\clearpage

\begin{thebibliography}{10}
\bibitem[Abbott et al.(2016)]{Abbott2016} Abbott, T., Abdalla, F.~B., Allam, S., et al.\ 2016, \prd, 94, 022001 
\bibitem[Blazek et al.(2011)]{Blazek2011} Blazek, J., McQuinn, M., \& Seljak, U.\ 2011,  JCAP, 5, 010
\bibitem[Blazek et al.(2015)]{Blazek2015} Blazek, J., Vlah, Z., \& Seljak, U.\ 2015,  JCAP, 8, 015
\bibitem[Blazek et al.(2017)]{Blazek2017} Blazek, J., MacCrann, N., Troxel, M.~A., \& Fang, X.\ 2017, arXiv:1708.09247
\bibitem[Bridle \& King(2007)]{Bridle2007} Bridle, S., \& King, L.\ 2007, New Journal of Physics, 9, 444 
\bibitem[Brown et al.(2002)]{Brown2002} Brown, M.~L., Taylor, A.~N., Hambly, N.~C., \& Dye, S.\ 2002, \mnras, 333, 501 
\bibitem[Catelan et al.(2001)]{Catelan2001} Catelan, P., Kamionkowski, M., \& Blandford, R.~D.\ 2001, \mnras, 320, L7 
\bibitem[Crittenden et al.(2001)]{Crittenden2001} Crittenden, R.~G., Natarajan, P., Pen, U.-L., \& Theuns, T.\ 2001, \apj, 559, 552 
\bibitem[Crittenden et al.(2002)]{Crittenden2002} Crittenden, R.~G., Natarajan, P., Pen, U.-L., \& Theuns, T.\ 2002, \apj, 568, 20 
\bibitem[Croft \& Metzler(2000)]{Croft2000} Croft, R.~A.~C., \& Metzler, C.~A.\ 2000, \apj, 545, 561
\bibitem[Heavens et al.(2000)]{Heavens2000} Heavens, A., Refregier, A., \& Heymans, C.\ 2000, \mnras, 319, 649
\bibitem[Heymans \& Heavens(2003)]{Heymans2003} Heymans, C., \& Heavens, A.\ 2003, \mnras, 339, 711 
\bibitem[Heymans et al.(2006)]{Heymans2006b} Heymans, C., White, M., Heavens, A., Vale, C., \& van Waerbeke, L.\ 2006, \mnras, 371, 750
\bibitem[Heymans et al.(2012)]{Heymans2012} Heymans, C., Van Waerbeke, L., Miller, L., et al.\ 2012, \mnras, 427, 146
\bibitem[Heymans et al.(2013)]{Heymans2013} Heymans, C., Grocutt, E., Heavens, A., et al.\ 2013, \mnras, 432, 2433  
\bibitem[Hilbert et al.(2017)]{Hilbert2017} Hilbert, S., Xu, D., Schneider, P., et al.\ 2017, \mnras, 468, 790 
\bibitem[Hildebrandt et al.(2016)]{Hildebrandt2016} Hildebrandt, H., Choi, A., Heymans, C., et al.\ 2016, \mnras, 463, 635 
\bibitem[Hildebrandt et al.(2017)]{Hildebrandt2017} Hildebrandt, H., Viola, M., Heymans, C., et al.\ 2017, \mnras, 465, 1454 
\bibitem[Hirata et al.(2004)]{Hirata2004a} Hirata, C.~M., Mandelbaum, R., Seljak, U., et al.\ 2004, \mnras, 353, 529
\bibitem[Hirata \& Seljak(2004)]{Hirata2004b} Hirata, C.~M., \& Seljak, U.\ 2004, \prd, 70, 063526 
\bibitem[Hirata et al.(2007)]{Hirata2007} Hirata, C.~M., Mandelbaum, R., Ishak, M., et al.\ 2007, \mnras, 381, 1197
\bibitem[Hui \& Zhang(2008)]{Hui2008} Hui, L., \& Zhang, J.\ 2008, \apj, 688, 742-756  
\bibitem[Jing(2002)]{Jing2002} Jing, Y.~P.\ 2002, \mnras, 335, L89 
\bibitem[Jing et al.(2007)]{Jing2007} Jing, Y.~P., Suto, Y., \& Mo, H.~J.\ 2007, \apj, 657, 664
\bibitem[Jing(2018)]{Jing2018} Jing, Y.~P., \ 2018, arxiv: 1807.06802
\bibitem[Joachimi \& Schneider(2008)]{Joachimi2008} Joachimi, B., \& Schneider, P.\ 2008, \aap, 488, 829 
\bibitem[Joachimi \& Schneider(2009)]{Joachimi2009} Joachimi, B., \& Schneider, P.\ 2009, \aap, 507, 105 
\bibitem[Joachimi et al.(2011)]{Joachimi2011} Joachimi, B., Mandelbaum, R., Abdalla, F.~B., \& Bridle, S.~L.\ 2011, \aap, 527, A26 
\bibitem[Joachimi et al.(2013)]{Joachimi2013} Joachimi, B., Semboloni, E., Hilbert, S., et al.\ 2013, \mnras, 436, 819 
\bibitem[Joudaki et al.(2017)]{Joudaki2017} Joudaki, S., Blake, C., Heymans, C., et al.\ 2017, \mnras, 465, 2033
\bibitem[King \& Schneider(2002)]{King2002} King, L., \& Schneider, P.\ 2002, \aap, 396, 411
\bibitem[King \& Schneider(2003)]{King2003} King, L.~J., \& Schneider, P.\ 2003, \aap, 398, 23  
\bibitem[King(2005)]{King2005} King, L.~J.\ 2005, \aap, 441, 47 
\bibitem[Kirk et al.(2010)]{Kirk2010} Kirk, D., Bridle, S., \& Schneider, M.\ 2010, \mnras, 408, 1502 
\bibitem[Lee \& Pen(2001)]{Lee2001} Lee, J., \& Pen, U.-L.\ 2001, \apj, 555, 106
\bibitem[Mandelbaum et al.(2006)]{Mandelbaum2006} Mandelbaum, R., Hirata, C.~M., Ishak, M., Seljak, U., \& Brinkmann, J.\ 2006, \mnras, 367, 611 
\bibitem[Mandelbaum et al.(2011)]{Mandelbaum2011} Mandelbaum, R., Blake, C., Bridle, S., et al.\ 2011, \mnras, 410, 844 
\bibitem[Munshi et al.(2008)]{Munshi2008} Munshi, D., Valageas, P., van Waerbeke, L., \& Heavens, A.\ 2008, \physrep, 462, 67
\bibitem[Okumura \& Jing(2009)]{Okumura2009a} Okumura, T., \& Jing, Y.~P.\ 2009, \apjl, 694, L83 
\bibitem[Okumura et al.(2009)]{Okumura2009b} Okumura, T., Jing, Y.~P., \& Li, C.\ 2009, \apj, 694, 214
\bibitem[Pereira et al.(2008)]{Pereira2008} Pereira, M.~J., Bryan, G.~L., \& Gill, S.~P.~D.\ 2008, \apj, 672, 825-833 
\bibitem[Schmidt et al.(2009)]{Schmidt2009} Schmidt, F., Rozo, E., Dodelson, S., Hui, L., \& Sheldon, E.\ 2009, \apj, 702, 593
\bibitem[Schneider \& Bridle(2010)]{Schneider2010} Schneider, M.~D., \& Bridle, S.\ 2010, \mnras, 402, 2127 
\bibitem[Singh et al.(2015)]{Singh2015} Singh, S., Mandelbaum, R., \& More, S.\ 2015, \mnras, 450, 2195 
\bibitem[Singh \& Mandelbaum(2016)]{Singh2016} Singh, S., \& Mandelbaum, R.\ 2016, \mnras, 457, 2301
\bibitem[Takada \& White(2004)]{Takada2004} Takada, M., \& White, M.\ 2004, \apjl, 601, L1
\bibitem[Troxel \& Ishak(2012a)]{Troxel2012a} Troxel, M.~A., \& Ishak, M.\ 2012a, \mnras, 419, 1804 
\bibitem[Troxel \& Ishak(2012b)]{Troxel2012b} Troxel, M.~A., \& Ishak, M.\ 2012b, \mnras, 423, 1663 
\bibitem[Troxel \& Ishak(2012c)]{Troxel2012c} Troxel, M.~A., \& Ishak, M.\ 2012c, \mnras, 427, 442 
\bibitem[Troxel \& Ishak(2015)]{Troxel2015} Troxel, M.~A., \& Ishak, M.\ 2015, \physrep, 558, 1 
\bibitem[Troxel et al.(2017)]{Troxel2017} Troxel, M.~A., MacCrann, N., Zuntz, J., et al.\ 2017, arXiv:1708.01538 
\bibitem[Tugendhat \& Sch{\"a}fer(2018)]{Tugendhat2018} Tugendhat, T.~M., \& Sch{\"a}fer, B.~M.\ 2018, \mnras, 476, 3460
\bibitem[van Uitert \& Joachimi(2017)]{Uitert2017} van Uitert, E., \& Joachimi, B.\ 2017, \mnras, 468, 4502
\bibitem[Wei et al.(2018)]{Wei2018} Wei, C., Li, G., Kang, X., et al.\ 2018, \apj, 853, 25
\bibitem[Xia et al.(2017)]{Xia2017} Xia, Q., Kang, X., Wang, P., et al.\ 2017, \apj, 848, 22
\bibitem[Yao et al.(2017)]{Yao2017} Yao, J., Ishak, M., Lin, W., \& Troxel, M.\ 2017,  JCAP, 10, 056  
\bibitem[Zhang(2010a)]{Zhang2010a} Zhang, P.\ 2010a, \apj, 720, 1090 
\bibitem[Zhang(2010b)]{Zhang2010b} Zhang, P.\ 2010b, \mnras, 406, L95 

\end{thebibliography}
\end{document}